\documentclass[reqno]{amsart}
\usepackage{amsaddr}
\usepackage[margin=1in]{geometry}
\usepackage{graphicx}
\usepackage{multicol}
\usepackage{hyperref}
\title[The probability distribution of scientific citations]{Nonuniversal power law scaling in the probability \\ distribution of scientific citations}
\author[G.J. Peterson, S. Press\'{e}, and K.A. Dill]{\small{George J. Peterson$^1$, Steve Press\'{e}$^2$, and Ken A. Dill$^{2}$}}
\address{$^1$Biophysics Graduate Group, $^2$Department of Pharmaceutical Chemistry \\ University of California -- San Francisco, San Francisco, CA 94158}
\email{\href{mailto:dill@maxwell.ucsf.edu}{dill@maxwell.ucsf.edu}}

\begin{document}

\begin{abstract}
We develop a model for the distribution of scientific citations.  The model involves a dual mechanism: in the \emph{direct mechanism}, the author of a new paper finds an old paper $A$ and cites it.  In the \emph{indirect mechanism}, the author of a new paper finds an old paper $A$ only \emph{via} the reference list of a newer intermediary paper $B$, which has previously cited $A$.  By comparison to citation databases, we find that papers having few citations are cited mainly by the direct mechanism.  Papers already having many citations (`classics') are cited mainly by the indirect mechanism.  The indirect mechanism gives a power-law tail.  The `tipping point' at which a paper becomes a classic is about 21 citations for papers published in the Institute for Scientific Information (ISI) Web of Science database in 1981, 29 for \emph{Physical Review D} papers published from 1975-1994, and 39 for all publications from a list of high $h$-index chemists assembled in 2007.  The power-law exponent is not universal.  Individuals who are highly cited have a systematically smaller exponent than individuals who are less cited. 
\end{abstract}

\maketitle
\thispagestyle{empty}

\begin{multicols}{2}

Commonly observed in nature and in the social sciences are probability distribution functions that appear to involve dual underlying mechanisms, with a `tipping point' between them.  Examples of such probability distributions include the distributions of city sizes \cite{Zipf_1949, Gabaix_1999}; fluctuations in stock market indices \cite{Gopikrishnan_1999, Plerou_1999}; U.S. firm sizes \cite{Okuyama_1999,Axtell_2001}; degrees of Internet nodes \cite{Holme_2007, Clauset_2009}; numbers of followers of religions \cite{Clauset_2009}; gamma-ray intensities of solar flares \cite{Newman_2005}; sightings of bird species \cite{Clauset_2009}; and citations of scientific papers \cite{Redner_1998, Newman_2001, Barabasi_2002, Redner_2004}.  In these situations, a distribution $p(k)$ may have exponential behavior for small $k$ and a power-law tail for large $k$.  Here we develop a generative model for one such dual-mechanism process, scientific citations, for which databases are large and readily available.  Here, $k$ represents the number of citations a paper receives, ranging from $0$ to hundreds or, sometimes, thousands.  $p(k)$ is the distribution of the relative numbers of such citations, taken over a database of papers.

There have been several important studies of power-law tails of distributions, including those involving scientific citations.  Price noted that highly cited scientific papers accumulate additional citations more quickly than papers that have fewer citations \cite{Price_1976}.  He called this `cumulative advantage' (CA): the probability that a paper receives a citation is proportional to the number of citations it already has.  Price showed that this rule asymptotically gives a power law for large $k$.  Power-law tails have been widely explored in various contexts and under different names -- `the rich get richer', the Yule process \cite{Yule_1925, Simon_1955}, the Matthew effect \cite{Merton_1968}, or preferential attachment \cite{Barabasi_1999}.  Barab\'{a}si and Albert noted that networks, such as the World Wide Web, often have power-law distributions of vertex connectivities, called `scale-free' behavior \cite{Barabasi_1999}.  Their model, called preferential attachment, leads to a fixed power-law exponent of $-3$.  Because many properties of physical systems near their critical points also display power-law behavior, and because such exponents are often \emph{universal} (\emph{i.e.,} independent of microscopic particulars of the system), it raises the question of which power-law distributions have universal exponents and which do not.

The tail of the scientific citations distribution has been fit by various distributions, including power law \cite{Redner_1998,Lehmann_2003}, log-normal \cite{Redner_2005}, and stretched exponential \cite{Laherrere_1998}.  Recently, Clauset, Shalizi, and Newman proposed detailed statistical tests for determining whether various data sets have true power-law tails \cite{Clauset_2009}.  In agreement with Redner's earlier analysis \cite{Redner_1998}, Clauset \emph{et al.} confirm that the 1981 data set studied by Redner is indeed well-fit by a power-law.

Our interest here is not just in the large-$k$ tails of such distribution functions.  We are interested also in the small-$k$ behavior and the tipping point between the two different regions.  After all, the preponderance of scientific papers are not cited very commonly.  Some previous models have explored both small-$k$ and large-$k$ regimes of citations.  In 2001, Krapivsky and Redner developed a rate equation method to obtain solutions for several generalizations of the CA model, including results for nonlinear connection probabilities \cite{Krapivsky_2001}.  Krapivsky and Redner proposed a `growing network with redirection' (GNR) for the citations network.  They proposed that new papers could randomly cite existing papers, or could be \emph{redirected} to one of the papers in its reference list.  The GNR mechanism leads to a distribution with a \emph{non-universal} scaling exponent, depending on the value of the redirection parameter.  An analysis of this mechanism for arbitrary out-degree distribution was carried out by Rozenfeld and ben-Avraham \cite{Rozenfeld_2004}.  Recently, Walker \emph{et al.} proposed a redirection algorithm to rank traffic to \emph{individual} papers, which, instead of an initial random attachment probability, used an exponentially decaying probability of citation, according to the age of the paper \cite{Walker_2007}.  There have been many variations proposed of the basic CA model, including CA with error tolerance \cite{Albert_2000}, with an attractiveness parameter \cite{Dorogovtsev_2000}, with a fitness parameter \cite{Bianconi_2001}, with memory effects \cite{Klemm_2002}, with hierarchical organization \cite{Ravasz_2003}, with aging nodes \cite{Hajra_2006}, and a number of others.  A useful overview of CA models, and power laws in general, is by Newman \cite{Newman_2005}.

Here, we develop a model to address three points of particular interest to us.  First, existing models focus on the power-law tail.  We are interested here in the full distribution function and the nature of the transition, or the `tipping point,' from one mechanism to the other.  Second, we seek a mechanism that illuminates why the `rich get richer' in scientific citations.  Third, a strictly linear attachment rule predicts a single fixed exponent, $\gamma = 3$, where $p(k) \propto k^{-\gamma}$.  Here, we ask whether the power-law exponent for scientific citations is a universal constant, as is often observed in the physics of critical phenomena, or whether the power-law exponent for citations is a non-universal parameter which varies from one dataset to another.

The two-mechanism model we propose here is similar to the GNR model studied in \cite{Krapivsky_2001}, generalized for an out-degree greater than one.  A general treatment of the GNR model with arbitrary out-degree distribution was given in \cite{Rozenfeld_2004}.  Here, we derive $p(k)$ explicitly for the specific case of a \emph{fixed} out-degree, and analyze the `tipping point' transition between the two mechanisms.  We then fit our $p(k)$ to several citations datasets, and examine how the interactions between the two mechanisms produces different distributions (with different tipping points) for each dataset.  By sorting our datasets according to $h$-index, we show that the scaling exponent, $\gamma$, \emph{decreases} systematically with increasing values of $h$.  We interpret the changes in the scaling exponent using a parameter of our model as an increasing bias towards \emph{indirect} citation of well-known scientists.

\section{A Two-Mechanism Model}

Consider a directed graph on which each node represents a scientific paper.  Each edge represents a citation of one paper by another.  An outgoing edge indicates \emph{giving} a citation, and an incoming edge indicates \emph{receiving} a citation.  At a given time, the graph has $N$ nodes, representing \emph{old} papers that are already part of the graph.  At each time step, a \emph{new} paper is published (a node is added to the graph).  Each new paper gives a fixed number of citations, $n$, distributed among the $N$ old papers.  Hence the total number of citations given is $Nn$, and the total number of citations received is also $Nn$.  In general, we consider situations in which $N$ is large.  Let $k$ be the number of incoming links (citations) that a paper has received.  For example, a paper that has received no citations from other papers has $k=0$.  Some `classic' papers have attracted more than $k=1000$ citations.   A given collection of papers will have a distribution, $p(k)$, of papers that have received $k=0, 1, 2, \ldots$ citations.

We first focus on a particular old paper, paper $A$.  The probability that a new paper will randomly link to paper $A$ is
\begin{equation}\label{eq:directcite}
r_{\text{direct}} = \frac{1}{N}.
\end{equation}
\noindent We call Equation~\ref{eq:directcite} the \emph{direct mechanism} of citations.\footnote{Because each new paper will not cite an old paper more than once, the direct probability, Eq. \ref{eq:directcite}, of the first citation is $1/N$, for the second citation is $1/(N-1)$, and so on, and for the $n^{\text{th}}$ citation is $1/(N-n+1)$.  For real-world graphs, however, $N$ is of the order of $500,000$ and $n$ is around $20$.  So, we assume $N \gg n$, and $1/(N-n+1) \sim 1/N$.  Similarly, the indirect probability, as $Nn \gg n$, Eq. \ref{eq:indirectcite} is approximately $k/(Nn-n+1) \sim k/(Nn)$.  Note also that, perhaps unrealistically, no special weight is given to the possibility of simultaneously citing both paper $A$ and one of its references.}

\begin{figure*}
\centerline{\includegraphics[width=1\textwidth]{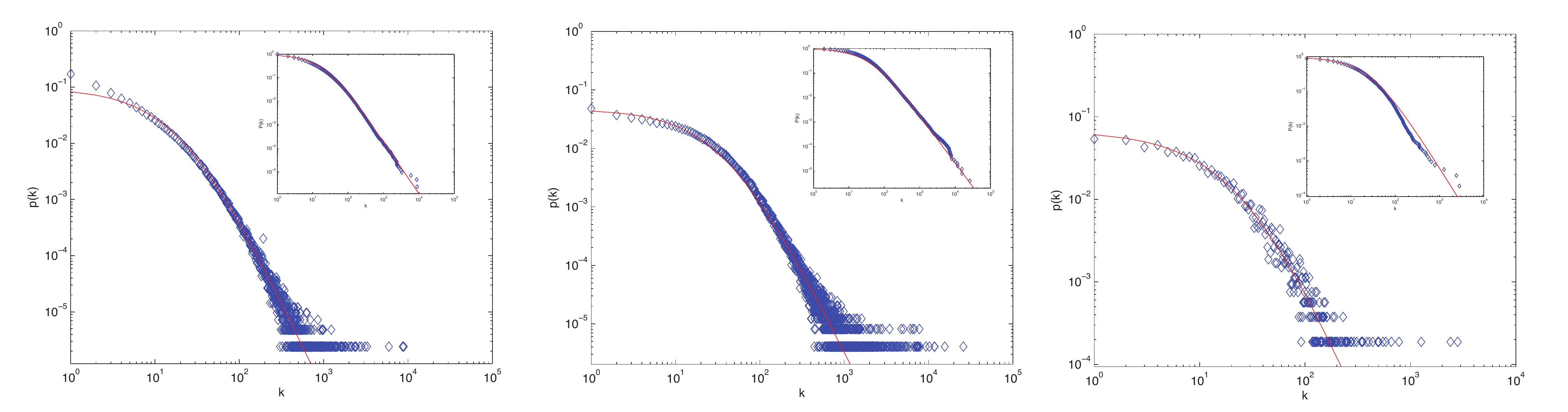}}
\caption{Probability of receiving exactly $k$ citations (PDF) and at least $k$ citations (CDF, inset) for datasets 1 (left), 2 (center), and 3 (right).  Empirical data points are shown as blue diamonds, and best-fit curves as solid red lines.}
\end{figure*}

In addition, scientific papers are also cited by an \emph{indirect mechanism}: the author of the new paper may first find a paper $B$ and learn of paper $A$ \emph{via} $B$'s reference list.  On the citation graph, searching through $B$'s reference list is a nearest-neighbor-link mechanism.  Suppose there are already $k$ incoming links to paper $A$.  Because there are a total of $nN$ incoming links to all papers, the probability that the author of the new paper randomly finds paper $A$, \emph{via} the reference list of some other paper is
\begin{equation}\label{eq:indirectcite}
r_{\text{indirect}} (k) = \frac{k}{Nn}.
\end{equation}

Given that the author of the new paper has found old paper $A$, the author will either cite a paper from $A$'s reference list with probability $c$, or cite $A$ itself with probability $1-c$.  If paper $A$ currently has $k$ citations, then the number of citations, $R(k)$, to paper $A$ from a new paper, through either the direct or indirect mechanism, is
\begin{eqnarray}\label{eq:pnet}
R(k) \!\!&=&\!\! n \left[\left(1-c\right) r_{\text{direct}} + c \, r_{\text{indirect}}(k) \right] \\
 \!\!&=&\!\! \frac{n(1-c)}{N} +\frac{kc}{N}.\notag
\end{eqnarray}

Next, we compute the in-link distribution $p(k)$, the fraction of the $N$ papers that have $k$ incoming citations.  The total number of papers having $k$ citations is $Np(k)$.\footnote{The in-link distribution should be considered a function of both $k$ and $N$, $p(k,N)$.  However, we find that in the large $N$ limit, the difference between $p(k,N)$ and $p(k,N-1)$ decreases as $1/N$.  It is therefore vanishingly small for very large $N$, and $\lim_{N\rightarrow \infty} p(k,N) = p(k)$.}  We calculate $p(k)$ using a difference equation to express the flows into and out of the bin of papers having $k$ citations for each time step (each time a new node is added).  The population of the bin of papers with $k$ citations increases every time a paper with $k-1$ citations receives another citation and decreases every time a paper that already has $k$ citations receives another citation,
\begin{align}\label{eq:mastereq}
p(k) \,\,=&\,\, N \left[ R(k-1) p(k-1) - R(k) p(k) \right] \\
=&\,\, \left[ n(1-c)+c(k-1) \right] p(k-1) - \notag\\
&\,\,\left[ n(1-c) + ck \right] p(k). \notag
\end{align}
Equation~\ref{eq:mastereq} rearranges to:
\begin{equation} \label{eq:itera}
p(k) = \frac{\alpha-1 +k}{\alpha+1/c+k} \cdot p(k-1).
\end{equation}
where, to simplify the notation, we have defined
\begin{equation}
\alpha = \frac{n}{c} -n.
\end{equation}

The equation for $p(0)$ involves no inflow from a lesser bin.  Instead, the inflow comes from the addition of a new paper per time step, which is 1 by definition.  The outflow term is calculated as for other values of $k$.  Therefore, $p(0) = 1 - n \left(1-c\right)p(0)$, which rearranges to:
\begin{equation}\label{eq:p0}
p(0) = \frac{1}{n-nc+1}.
\end{equation}
\noindent Substituting in Equation~\ref{eq:p0} and applying Equation~\ref{eq:itera} recursively gives\footnote{The factorials in Equation~\ref{eq:pfactorial} are understood to be gamma functions for non-integer $1/c$ values.  To show that equation \ref{eq:pfactorial} is normalized, we use \[ \sum_{k=0}^{\infty} \frac{(\alpha-1+k)!}{(\alpha+1/c+k)!} = \left(\alpha c + 1\right) \frac{(\alpha-1)!}{(\alpha+1/c)!}. \] Substituting into \ref{eq:pfactorial}, we find that $\sum_k p(k) = 1$, as required.}  
\begin{equation}\label{eq:pfactorial}
p(k) = \frac{1}{\alpha c+1} \cdot \frac{(\alpha-1+k)! (\alpha+1/c)!}{(\alpha-1)! (\alpha+1/c+k)!}.
\end{equation}
\end{multicols}
\begin{table*}[htdp]
\caption{Fitting parameters for datasets 1-3}
\begin{center}
\begin{tabular*}{\hsize}{@{\extracolsep{\fill}}l l l l l l}
Dataset & $c$ & $n$ & $\gamma$ & $\alpha$ & $N$\\ 
\hline
1. All 1981 publications & $0.454 \pm 0.004$ & $17.3 \pm 0.3$ & $3.20 \pm 0.02$ & $20.8 \pm 0.4$ & $415229$\\
2. High $h$-index chemists & $0.517 \pm 0.001$ & $42.0 \pm 0.1$ & $2.935 \pm 0.005$ & $39.2 \pm 0.1$ & $245461$\\
3. \textit{Phys. Rev. D} publications & $0.48 \pm 0.03$ & $27 \pm 2$ & $3.1 \pm 0.1$ & $29 \pm 3$ & $5327$\\
\hline
\end{tabular*}
\end{center}
\label{default}
\end{table*}
\begin{multicols}{2}
\noindent When $\alpha$ is sufficiently large, we apply Stirling's approximation to Equation~\ref{eq:pfactorial}, which yields
\begin{eqnarray}\label{eq:pstirling}
p(k) \approx \frac{(\alpha+1/c)^{\alpha+1/c}}{(\alpha c + 1) \left(\alpha-1\right)^{\alpha-1}} \left( \frac{\alpha-1+k}{\alpha+1/c+k} \right)^{\alpha+k} \notag\\
\times \left(\alpha-1+k\right)^{-1} \left( \alpha+\frac{1}{c}+k \right)^{-1/c}.
\end{eqnarray}
In the large-$k$ tail ($k \gg \alpha$), we have
\begin{equation*}
\left( \frac{\alpha-1+k}{\alpha+1/c+k} \right)^{\alpha+k} \approx e^{-(1+1/c)},
\end{equation*}
and
\begin{equation*}
(\alpha-1+k)^{-1} \left(\alpha+\frac{1}{c}+k\right)^{-1/c} \approx k^{-(1+1/c)}.
\end{equation*}
Therefore, Equation~\ref{eq:pstirling} becomes, in the large-$k$ tail:
\begin{equation}\label{eq:pscaling}
p(k) \approx  \left[ \frac{(\alpha+1/c)^{\alpha+1/c}e^{-(1+1/c)}}{(\alpha c + 1) \left(\alpha-1\right)^{\alpha-1}} \right] k^{-\left(1+1/c \right)}.
\end{equation}

Equation \ref{eq:pstirling} gives our model's prediction for the distribution of citations.  It expresses both the direct and indirect citation mechanisms.  Equation \ref{eq:pscaling} indicates that once a paper's number of citations, $k$, is large enough, further citations of that paper undergo a sort of runaway growth because there are so many ways to find it through other papers that have already cited it; for scientific citations, `the rich get richer.'  The `tipping point' where $r_{\text{indirect}}$ overtakes $r_{\text{direct}}$ happens at
\begin{equation}\label{eq:ktip}
k = \alpha.
\end{equation}
For example, if $c=1/2$ and the average paper in the database gives out $n = 15$ citations, then after any particular paper in that database has received 15 citations, it will begin to accumulate citations significantly faster than random -- it will have `tipped over' into the power-law scaling region.  In this region, the power law exponent,
\begin{equation}
\gamma = 1+\frac{1}{c},
\end{equation}
is determined by the parameter $c$.  Hence, `cumulative advantage' arises in our model because there are more routes (through the reference lists of other papers) for finding a classic paper than for finding a non-classic paper.

\section{The Datasets}

Figure 1 shows fits to normalized empirical probability distribution functions (PDFs, the probability of receiving \emph{exactly} $k$ citations) and complementary cumulative distribution functions (CDFs, the probability of receiving \emph{at least} $k$ citations), $P(k) = \int_{k'}^\infty p(k') dk'$ , for three datasets:

\begin{enumerate}
\item Citations of publications catalogued in the ISI Web of Science database in 1981 \cite{Redner_1998}
\item Citations of publications by authors on a 2007 list of the living highest $h$-index chemists \cite{Peterson_2007}
\item Citations of publications in the \textit{Physical Review D} journal from 1975-1994 \cite{Redner_1998}
\end{enumerate}

\noindent Datasets 1 and 3 were downloaded from Sidney Redner's website\footnote{http://physics.bu.edu/$\sim$redner/projects/citation/index.html}.  We gathered dataset 2 from the ISI Web of Knowledge\footnote{http://isiwebofknowledge.com} using a Python script.  Parameters for these fits are shown in Table 1, and plots of the datasets and best-fit $p(k)$ distributions are shown in Figure 1.  We also sorted dataset 2 by $h$-index.  Parameters for different $h$-index ranges are shown in Table 2, and fits are shown in Figure 2.  The relation between our estimates of $\gamma$ and $h$ is shown in Figure 3.  To obtain estimates and 95\% confidence intervals of $c$ and $n$, we used Matlab's implementation of the iteratively reweighted least squares algorithm, using bisquare weights \cite{Mosteller_1977}.  All curve fitting was applied to the raw (not binned or log-transformed) data.

\begin{figure*}
\centerline{\includegraphics[width=0.55\textwidth]{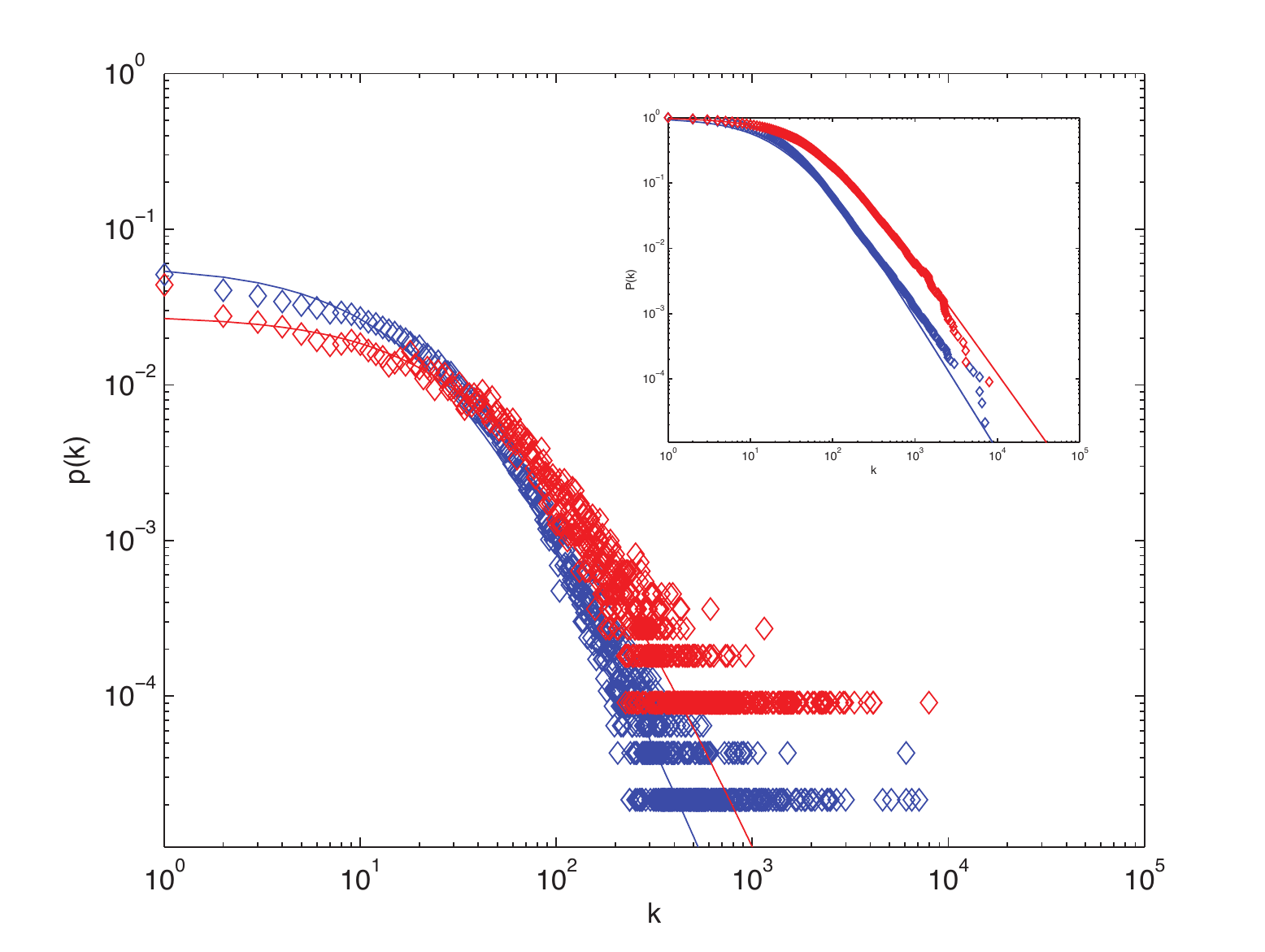}}
\caption{Comparison of the normalized PDFs and CDFs (inset) for chemists with $h = $ 100+ (red) and chemists with $h = $ 50-53 (blue).}
\end{figure*}

\section{Results}

Our model has two parameters: $n$, the average number of citations given out by all the papers in the database, and $c$, the chance of citing from a paper's reference list.   The model power-law exponent is then fixed by the relationship $\gamma = 1 + 1/c$.  Our best fit of dataset 1 gives a value of $n = 17.3 \pm 0.3$, in approximate agreement with the independent estimate of $15.01$ found for papers published in 1980 \cite{Biglu_2007}.  Also, our predicted value of $\gamma = 3.20 \pm 0.02$ agrees with the best-fit power-law exponent previously found by Clauset, of $\gamma = 3.16$ \cite{Clauset_2009}.  Table 1 shows the best-fit parameter values for the three different datasets.

We explored the $p(k)$ distributions for small groups of scientists, as shown in Figure 2.  We wanted to test an alternate hypothesis that some scientists might publish only low-$k$ papers and others might publish only classic high-$k$ papers.  Our limited tests argue against this hypothesis.  Figure 2 indicates that even highly cited scientists have more low-$k$ papers than high-$k$ papers.  One reason is that every publication in the scientific literature is new for a while, and requires some time to become highly cited.

Interestingly, the slope of the power-law region differs between the two groups shown in Figure 2.  To examine this difference in more detail, we parsed dataset 2 by $h$-index (Table 2).  The $h$-index of a scientist is defined as the point where $h$ of the scientist's papers have at least $h$ citations each \cite{Hirsch_2005}.  That is, $h$ is defined by the requirement to satisfy the expression, $N p(h) = h$.  There is no simple analytical relationship between a scientist's $h$-index and the parameters of our model.

From Table 2, we conclude that $c$ increases with $h$-index, indicating that there is a bias towards selecting papers out of a reference list that were written by scientists who are already very highly cited (Figure 2).  This bias may reflect the tendency of authors who, scanning a paper's references for further information, are more likely to select a paper written by an author they have previously heard of.  The more highly cited the scientist, the lower his or her power-law exponent (\emph{i.e.}, the fatter the tail); see Figure 3.  The error bars are sufficiently small to indicate that these trends are real, and that there is not a single universal exponent, such as $\gamma = 3$; rather, the exponent depends on the subset of scientists examined.  Note that, here, we consider a scientist to have authored a paper if his or her name appears anywhere in the list of authors.  An interesting question for future work might be to examine whether this effect is changed by only considering the $h$-index of each paper's leading and/or corresponding author.

Our model bears some resemblance to Price's application of CA to scientific citations \cite{Price_1976}.  One key difference is that our two parameters both have physical meaning.  To avoid the issue of new papers having a citation probability of zero when $k=0$, Price proposed that the citation probability should be proportional instead to $k+w$, where $w$ is a constant that he refers to as a `fudge factor.'  He sets $w=1$, although as later noted by Newman, there does not seem to be a good reason to choose this value \cite{Newman_2005}.  The connection rule for our model is given by Equation \ref{eq:pnet}, and suggests a simple interpretation: Price's constant arises from random connections, and the tipping point, Equation \ref{eq:ktip}, is determined by the average size of the reference lists given out per paper, and the probability of searching through those reference lists.

This two-mechanism model also provides a justification for a CA mechanism.  Barab\'{a}si and Albert remarked that CA only produced a power law distribution when the connection probability was linearly proportional to $k$ \cite{Barabasi_1999}, but it was not clear what was special about linearity.  The present model presents a possible explanation for the existence of this mechanism, and why the $k$ dependence should be linear: $k$ appears in $r_{\text{indirect}}$ because a paper's $k$ incoming citations are represented by $k$ nearest-neighbor links on the graph.

\end{multicols}
\begin{figure*}
\centerline{\includegraphics[width=0.55\textwidth]{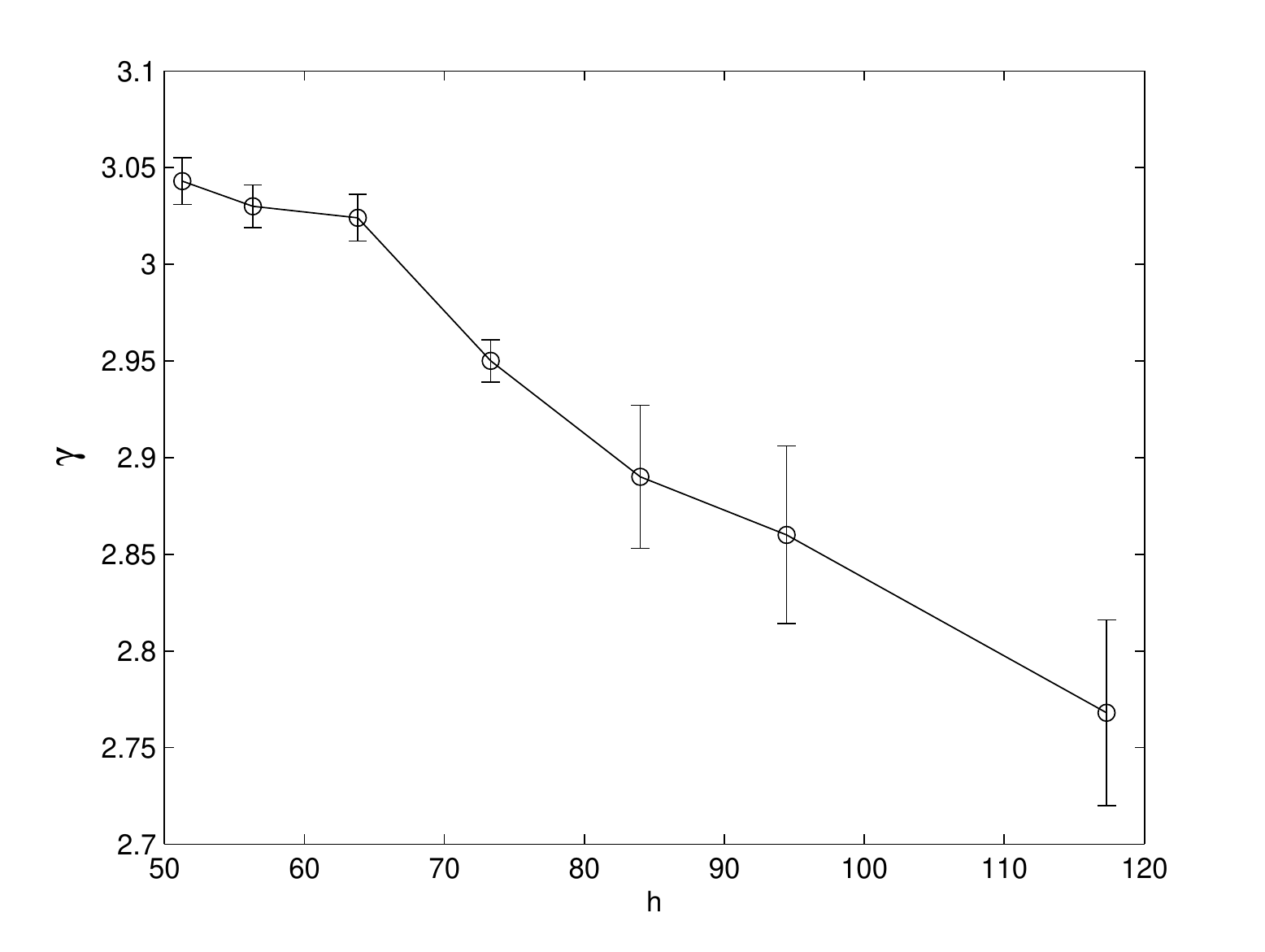}}
\caption{Power-law exponent $\gamma$ plotted against $h$-index for subsets of dataset 2.}
\end{figure*}
\begin{multicols}{2}

\section{Conclusion}

We have developed a model of scientific citations, involving both direct and indirect routes to finding and citing papers.  This two-mechanism model predicts exponential behavior in the small-$k$ region and power law tails in the large-$k$ region.  One parameter of the model, $n$, is the average number of citations given out per paper.  Our best-fit value of $n$ is consistent with an independent, empirical measure of it made by Biglu \cite{Biglu_2007}.  Our other parameter, $c$, defines the power-law exponent, $\gamma = 1 + 1/c$, which is in agreement with data previously evaluated in \cite{Clauset_2009}.  Two key findings here are:  (1) the tipping point for a paper to reach `classic-paper' status, \emph{i.e.} its power-law citation region, is about 21 citations for the ISI Web of Science database, and (2) the power-law exponent is not a universal feature of all scientific citations.  The exponent diminishes systematically with increasing $h$-index of a scientist.  Our model describes systems that are governed by random choices in the small-$k$ region, cumulative advantage in the high-$k$ region, and a tipping point between them.

\section*{acknowledgments}
We thank A\'{e}thalie Chabriol for assistance with data acquisition, Kristin Peterson for helpful discussions of curve-fitting methods, and Aaron Clauset, Kingshuk Ghosh, Sergei Maslov, Mark Newman, and Sid Redner for feedback on the manuscript.  GJP is grateful for financial support from an NDSEG Fellowship from the Department of Defense, SP thanks the FQRNT, and KD and SP appreciate the support from NIH GM 34993.  We thank the ISI Web of Science for their permission to use this data, and Sid Redner for providing a publicly available database of citations.

\end{multicols}
\begin{table}[htdp]
\caption{Fitting parameters for $h$-index ranges within dataset 2}
\begin{center}
\begin{tabular*}{\hsize}{@{\extracolsep{\fill}}c l l l l l}
$h$ range & $c$ & $n$ & $\gamma$ & $\alpha$ & $N$\\ 
\hline
100+ & $0.57 \pm 0.01$ & $80 \pm 3$ & $2.77 \pm 0.05$ & $60 \pm 2$ & $11029$\\
90-99 & $0.54 \pm 0.01$ & $77 \pm 3$ & $2.86 \pm 0.05$ & $66 \pm 3$ & $11476$\\
80-89 & $0.53 \pm 0.01$ & $60 \pm 2$ & $2.89 \pm 0.04$ & $53 \pm 2$ & $15408$\\
70-79 & $0.513 \pm 0.003$ & $40.6 \pm 0.4$ & $2.95 \pm 0.01$ & $38.5 \pm 0.4$ & $54236$\\
60-69 & $0.494 \pm 0.002$ & $48.7 \pm 0.4$ & $3.02 \pm 0.01$ & $49.9 \pm 0.5$ & $56052$\\
54-59 & $0.493 \pm 0.003$ & $34.9 \pm 0.3$ & $3.03 \pm 0.01$ & $35.9 \pm 0.4$ & $44715$\\
50-53 & $0.489 \pm 0.003$ & $31.3 \pm 0.3$ & $3.04 \pm 0.01$ & $32.7 \pm 0.4$ & $46421$\\
\hline
\end{tabular*}
\end{center}
\label{default}
\end{table}
\begin{multicols}{2}

\end{multicols}

\end{document}